\documentclass[preprint2]{aastex}%

\usepackage{epsfig}
\usepackage{graphicx}

\shorttitle{CLOWFS for high performance Coronagraphs}
\shortauthors{Vogt, Martinache, Guyon et al.}

\begin{document}

\title{Coronagraphic Low Order Wave Front Sensor : post-processing
  sensitivity enhancer for high performance coronagraphs}

\author{Fr\'ed\'eric P. A. Vogt\altaffilmark{1}, Frantz Martinache,
  Olivier Guyon, Takashi Yoshikawa, Kaito Yokochi, Vincent Garrel and
  Taro Matsuo}
\affil{National Astronomical Observatory of Japan, Subaru Telescope,
  Hilo, HI 96720, USA}
\email{frantz@naoj.org}

\altaffiltext{1}{Now at : Mount Stromlo and Siding Spring
  Observatories, Research School of Astronomy and Astrophysics,
  The Australian National University, Cotter Road, Weston Creek, ACT
  2611, Australia.}

\begin{abstract}
Detection and characterization of exoplanets by direct imaging
requires a coronagraph designed to deliver high contrast at small
angular separation. To achieve this, an accurate control of low order
aberrations, such as pointing and focus errors, is essential to
optimize coronagraphic rejection and avoid the possible confusion
between exoplanet light and coronagraphic leaks in the science image.
Simulations and laboratory prototyping have shown that a Coronagraphic
Low Order Wave-Front Sensor (CLOWFS), using a single defocused image
of a reflective focal plane ring, can be used to control tip-tilt to
an accuracy of $10^{-3}$ $\lambda{/D}$. This paper demonstrates that
the data acquired by CLOWFS can also be used in post-processing to
calibrate residual coronagraphic leaks from the science image.
Using both the CLOWFS camera and the science camera in the system, we
quantify the accuracy of the method and its ability to successfully
remove light due to low order errors from the science image. We also
report the implementation and performance of the CLOWFS on the Subaru
Coronagraphic Extreme AO (SCExAO) system and its expected on-sky
performance.
In the laboratory, with a level of disturbance similar to what is
encountered in a post Adaptive Optics beam, CLOWFS post-processing has
achieved speckle calibration to 1/300 of the raw speckle level. This
is about 40 times better than could be done with an idealized PSF
subtraction that does not rely on CLOWFS.
\end{abstract}

\keywords{Extrasolar Planets --- Astronomical Instrumentation}

\section{Introduction}

Since the discovery of 51 Pegasi by \citet{1995Natur.378..355M},
several hundreds extrasolar planets have been uncovered, for the most
part via indirect detection methods such as radial velocity
measurements of the reflex motion of their host star and photometric
transit. Direct imaging was recently able to produce the first high
contrast images of  \emph{solid} planetary candidates orbiting nearby
main sequence stars: Fomalhaut \citep{2008Sci...322.1345K}, Beta
Pictoris \citep{2009A&A...493L..21L} and HR 8799
\citep{2008Sci...322.1348M, 2009ApJ...694L.148L}. Following up on
these direct detections, \citet{2010ApJ...710L..35J} obtained the
first direct extrasolar planet spectrum, opening the way to better
characterization of planetary atmospheres. These direct detections are
currently limited to planets at large orbital separations of several
dozens of AU, and only probe a small fraction of the distribution of
known extrasolar planets, for which  the median orbital separation is
likely closer to 1 AU.

At near infrared wavelength, an 8-meter class telescope provides
sufficient angular resolution (40 milli-arcseconds at
$\lambda$=$\sim$1.6 ${\mu}$m) to be able to detect companions in the
Habitable Zone of nearby stars (d $<$ 30 pc). High contrast imaging
near the diffraction limit however requires very good control and
calibration of the wavefront aberrations in the optical system. From
the ground, this task is complicated by the rapidly changing
atmospheric wavefront creating speckled images. While Adaptive Optics
(AO) offers a major improvement, performance is still limited and
direct imaging of planets often requires post-processing techniques
such as Angular Differential Imaging (ADI)
\citep{2006ApJ...641..556M}, which is most efficient at large
($\gtrapprox$ 10 $\lambda/D$) angular separations.

In the near future, high contrast imaging systems employing new
coronagraphic and wavefront control techniques will greatly improve
our ability to image exoplanets. These Extreme-AO systems include the
Gemini Planet Imager (GPI) \citep{2006SPIE.6272E..18M}, ESO's SPHERE
\citep{2008SPIE.7014E..41B} and Subaru's SCExAO
\citep{2009SPIE.7440E..20M}. They will use efficient coronagraphs and
high-speed high-order wavefront corrections to reduce speckles in the
coronagraphic image. The images of the HR 8799 planetary system
obtained by \citet{2010AAS...21537706S} on a well-corrected 1.5 meter
aperture using a high-performance coronagraph demonstrate the
relevance of this approach. 

Imaging companions close to the edge of the occulting mask in a
coronagraph (that is $\sim$40 mas on SCExAO) however remains an
unachieved feat. The current state of the art for PSF calibration is
an optimized version of ADI called LOCI introduced by
\citet{2007ApJ...660..770L}.
LOCI can calibrate out high-order aberrations to a very high level of
contrast (12 magnitudes and higher), but like any technique relying on
ADI, only for angular separations greater than 10 $\lambda/D$.
Detection at the edge of a $\sim$ 1 $\lambda/D$ occulter is extremely
sensitive to low-order aberrations such as pointing and focus.

Of these aberrations, pointing is especially critical, since near the
occulter, a tip-tilt excursion along a given direction mimics the
signal of a true companion in a coronagraphic image.
This issue, first identified for high contrast space borne coronagraphs,
has been addressed by \cite{2009ApJ...693...75G}, with the
Coronagraphic Low-Order Wavefront Sensor (CLOWFS), a scheme using the
light occulted by a modified focal plane mask as an accurate pointing
tracker.

The idea of using the light otherwise lost in the coronagraphic focal
plane for tracking is not a novelty. It was for instance successfully
implemented on the LYOT project \citep{2006ApJ...650..484D}. The
calibration unit of GPI also uses the light occulted by the focal
plane mask to measure low order aberrations, after re-imaging the
pupil in a Shack Hartmann type wavefront sensor 
\citep{2010SPIE.7736E.179W}: pointing performance with this scheme
reaches 2 mas for typical expected observing conditions.

Maximum sensitivity to pointing errors is reached when the light from
opposite edges of the pupil is allowed to interfere, which naturally
happens in the focal plane. In this respect, while robust to a wide
range of errors, a Shack-Hartman doesn't appear optimal to measure
pointing: because it splits the pupil into sub-pupils, this capability
is lost, resulting in lesser performance than a focal-plane based
wavefront sensing scheme \citep{2005ApJ...629..592G}.

The originality of the CLOWFS design resides in its dual-zone focal
plane mask, designed to suppress a strong offset to the signal,
carrying most of the power but no information, in a manner reminiscent
of strioscopy. The suppression of this offset turns the otherwise
imperceptible changes due to small pointing errors into a macroscopic
change of the CLOWFS image. Using this scheme,
\citet{2009ApJ...693...75G} were able to stabilize tip-tilt at the
level of $10^{-3} \lambda/D$ in a closed-loop system for $\lambda=633$
nm, in a laboratory experiment.
The current implementation of CLOWFS on SCExAO exhibits pointing
residuals $<$0.2 mas, with a 50 Hz frame rate.

While this level of performance is quite remarkable, we demonstrate in
this work that additional calibration can be achieved in
post-processing, and lead to an improved subtraction of coronagraphic
leaks due to low-order aberrations in a long exposure.
Using the SCExAO system as a testbed, we experimentally demonstrate a
40 times improvement of the detection limit over a classical
calibration procedure at angular separation of a few $\lambda/D$.

This paper is organized as follows : in Section~\ref{sec:method}, we
introduce how to use CLOWFS for post processing of coronagraphic
images. In Section~\ref{sec:exp}, we describe the implementation of
the concept on the SCExAO experiment of the Subaru telescope, and we
present our results in Section~\ref{Sec:results}. In
Section~\ref{sec:discussion}, we summarise our results and discuss
possible updates to the presented CLOWFS configuration.

\begin{figure*}[htb!]
\centerline{\includegraphics[scale=0.4]{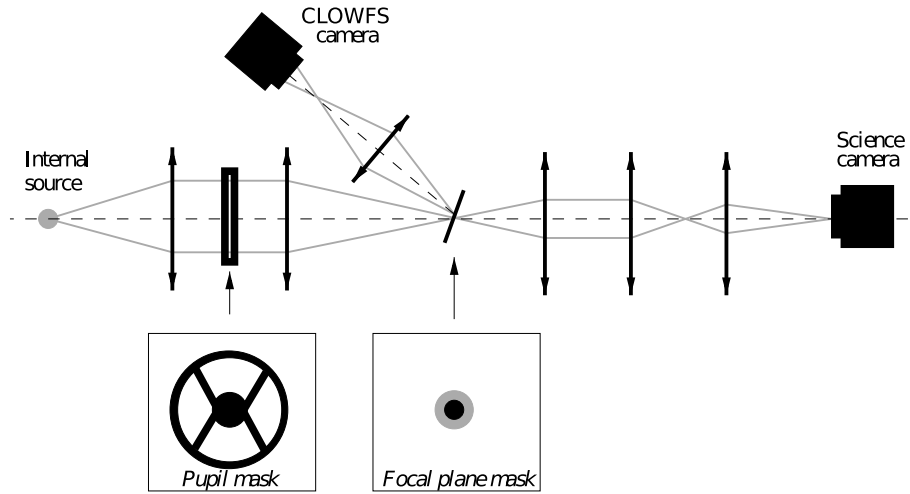}}
\caption{Optical layout of SCExAO used in Lyot-coronagraph mode. A
  mask lit by an IR ($\lambda=1.55$ ${\mu}m$) laser diode emulates the
  Subaru Telescope pupil. The beam is focused on the dual-zone
  focal-plane occulting mask described in the text. The light
  reflected by the focal-plane mask feeds the (slightly defocused)
  CLOWFS detector, used to characterize pointing errors. The light
  that is not intercepted by the mask is then re-imaged on the
  "science" detector. Examples of images obtained on both cameras are
  presented in Fig.~\ref{fig:clowfs_img} and \ref{fig:sci_img}.}
\label{fig:bench} 
\end{figure*}

\section{CLOWFS post-processing principle}
\label{sec:method}

A high performance PSF calibration procedure such as LOCI
\citep{2007ApJ...660..770L}, is based on a direct analysis of the
science data alone. Yet near the edge of the coronagraphic mask, it
is impossible, from such data only, to distinguish between the
signal of an actual faint companion and the one of a systematic
tip-tilt excursion off the coronagraphic mask at a given azimuth, that
would be for instance due to a vibration.
Data acquired with CLOWFS during a long exposure on the science camera
can however be used to discriminate the two situations, in
post-processing.

The scheme proposed in this paper is a form of adaptive optics PSF
reconstruction, which uses measurements acquired in a wavefront sensor
to estimate the long-exposure PSF in the science camera
\citep{1997JOSAA..14.3057V, 2006A&A...457..359G}. Adaptive Optics PSF
reconstruction has been implemented on several adaptive optics systems
\citep{2000A&AS..142..119H,2010SPIE.7736E..48J}, and relies on the
fact that the wavefront sensor, by measuring residual wavefront errors
at a fixed spatial sampling, can be used to estimate the inner part of
the PSF in the science camera. This estimation can be done at the
wavefront sensing sampling speed (typically 100 Hz to 1kHz), and is
then averaged for the duration of the science exposure.
In this paper, we reconstruct the very inner part of the coronagraphic
PSF using CLOWFS telemetry. Compared to previous implementations of
adaptive optics PSF reconstruction, our scheme is better suited to
high contrast coronagraphic imaging, as it uses a sensor which is
highly sensitive to low order aberrations and free of non-common path
errors. Our PSF reconstruction is however limited to the very inner
part of the coronagraphic PSF, and is most effective in the 1
$\lambda$/D wide area immediately around the focal plane mask. At
larger angular separation, wavefront errors can create speckles
without producing a signal in the CLOWFS camera, and other calibration
approaches must be used to reconstruct the PSF: for example telemetry
from a higher order WFS, differential spectral imaging, or the
ADI/LOCI technique. We also propose to use a new empirical image-based
PSF reconstruction algorithm, where CLOWFS images are matched to
science camera images to reconstruct the PSF, as opposed to relying on
a model of the adaptive optics system. This empirical approach is more
robust, simple to implement, and is made possible in our case by the
small number of modes measured by the CLOWFS.

A thorough description of the theory and hardware implementation of
CLOWFS was provided by \citet{2009ApJ...693...75G}. Here, it suffices
to remember that it operates thanks to an optimized dual zone focal
plane mask, absorbing at its center, and reflective in an annulus
whose outer edge defines the inner working angle of the system (1.5
$\lambda/D$ for SCExAO).
Figure \ref{fig:bench} shows the actual implementation of CLOWFS on
SCExAO. A lens re-images the reflective ring of the occulting mask on
a detector deliberately placed out of focus, referred to as the CLOWFS
camera.

\begin{figure}[htb!]
\centerline{\includegraphics[scale=0.3]{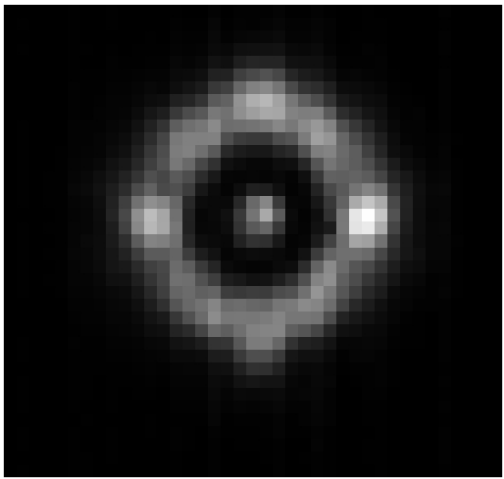}}
\caption{Example of CLOWFS camera image. Variations in the
  distribution of intensity in the image of the reflective annular
  region of the mask are used to track minute changes in the
  pointing.}
\label{fig:clowfs_img}
\end{figure}

During the (long) exposure on the science camera, the CLOWFS camera
acquires a continuous stream of short (typically millisecond)
exposures. An example of one such CLOWFS camera image is provided in
Fig. \ref{fig:clowfs_img}: variations of the distribution of intensity
in this image are used to identify drifts in pointing as well as
changes in focus.

\citet{2009ApJ...693...75G} have demonstrated that over a
small range ($\sim 0.2 \lambda/D$) of pointing errors, a linear model
relates the changes in CLOWFS images to the actual pointing error, and
took advantage of this in a close-loop system, stabilizing the
pointing at the level of $10^{-3} \lambda/D$ over extended periods of
time ($\sim 1$ hr).

Additional calibration of the coronagraphic leaks due to the residual
tip-tilt error can be achieved in post-processing. This calibration
relies on the one-to-one correspondence that exists between images
simultaneously acquired on the science and the CLOWFS
cameras. In either close or open loop, recording at high temporal rate
CLOWFS images during a long exposure on the science camera can help
predict the level of coronagraphic leaks attributable to pointing
errors. A synthetic pointing leak image can be built and subtracted
from the science image, to calibrate out the low-order aberrations
residuals, ultimately improving contrast limits.

\begin{figure*}[htb!]
\centerline{\includegraphics[scale=0.5]{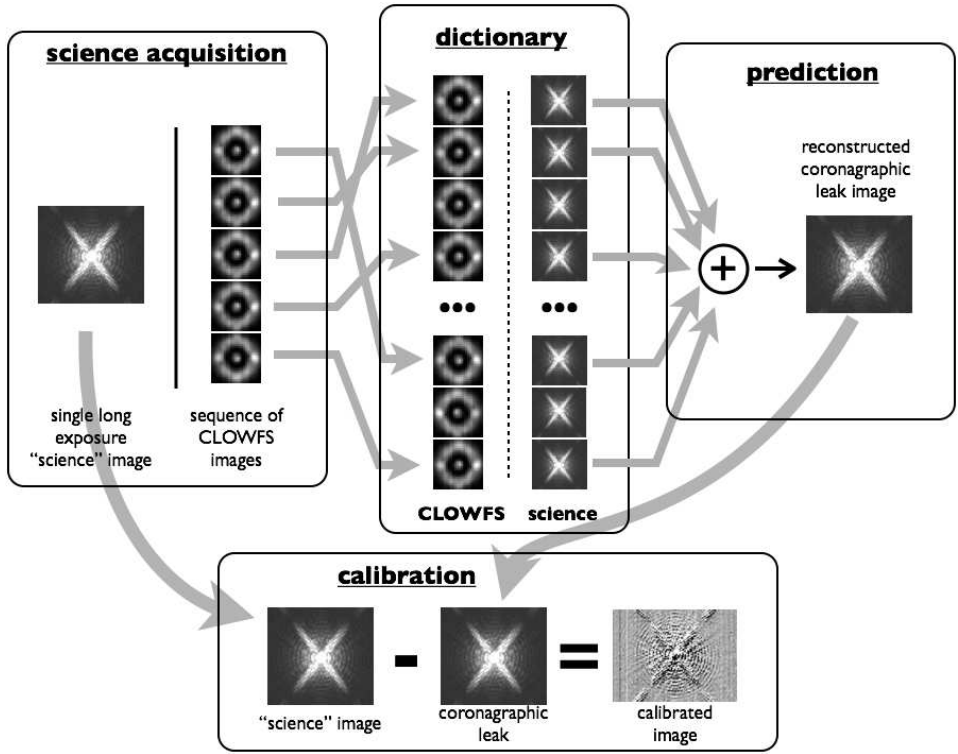}}
\caption{CLOWFS post-acquisition pointing errors calibration
  procedure, defined as the \emph{Match-Maker} Algorithm
  (MMA).}\label{fig:mma}
\end{figure*}

In its simplest version (see Section \ref{sec:discussion} for a
discussion of the possible complements), the post-acquisition
calibration of pointing errors with CLOWFS is a three-step procedure,
described in the following sections and illustrated in
Fig. \ref{fig:mma}.

\begin{itemize}
\item {Step 1:} calibrate the static optical configuration of the
  system (i.e. fine variations in optical path and alignment), using
  simultaneous pairs of images acquired with the science and the
  CLOWFS camera. This bank of pairs of images will be referred to as
  the dictionary (see Section~\ref{Sec:build}).
\item {Step 2:} track low-order modes during a long science exposure,
  continuously acquiring images at fast frame rate with the CLOWFS
  camera (see Section~\ref{Sec:compsub}).
\item {Step 3:} post processing of the data, by identifying a match
  for each CLOWFS image acquired during the science exposure in the
  dictionary (see Section~\ref{Sec:compsub}).
\end{itemize}

\subsection{System calibration: building up the dictionary}
\label{Sec:build}

During step 1, pairs of simultaneous short exposure images are
acquired with both the CLOWFS on a calibration source (internal source
or single star). Images for exposure \#$k$ are respectively labeled
DC$_{k}$ and DS$_{k}$ for the CLOWFS and science image. These images
are stored into a database, called the dictionary.

After one such acquisition, as long as the optics remain stable, the
CLOWFS image can be used as a key that points to a coronagraphic leak
term (the corresponding science image).
Later, any instant CLOWFS image can be compared to the entries in the
dictionary: the best matching entry in the dictionary allows to
predict the amount of coronagraphic leak.

While slow varying non-common path errors can occur after the
coronagraph, their effect can be kept small by minimizing the number
of optics used after the CLOWFS focal-plane mask, as well as by
regularly refreshing the content of the dictionary, so as to keep it
up to date with the current status of the system.

A preliminary version of the dictionary can therefore be compiled in
the lab prior to the actual observing, using a calibration
source. However to minimize systematic error terms, it must be
completed with more up-to-date images acquired on a series of
calibration (non-resolved) stars of spectral type and magnitude
comparable to the science target, so as to get the best possible
match. Ideally, during acquisition on the calibration star, one wants
to cover a range of pointing errors that is larger than experienced
during the science exposure.

It is important to emphasize here that low-order aberrations errors
are not explicitly calculated: instead, their consequences on the
coronagraphic image are directly recorded, via the CLOWFS system. 
Pragmatic, this approach eliminates the need for a high fidelity (yet
most likely imperfect) model of the coronagraph: ultimately, the
ability to precisely characterize the coronagraphic leaks is
determined by the coverage of the dictionary which can be made
arbitrarily large.

\subsection{Computing and subtracting coronagraphic leaks}
\label{Sec:compsub}

During a long (i.e. typically greater than one second) exposure on the
science camera, the CLOWFS camera acquires a sequence of short
exposures (Step 2), labeled C$_{i}$ that are saved along with the
science camera image (cf. Fig.~\ref{fig:mma}).
To predict the coronagraphic leak in the science image, a matching
algorithm (referred to as the \textit{Match-Maker Algorithm}, or MMA)
identifies each image of the CLOWFS sequence $C_i$ to the CLOWFS image
of the dictionary $DC_k$ that best matches it.
Fig. \ref{fig:clowfs_img} shows the typical structure of one CLOWFS
image. The region of interest within the image is defined as a disk
circumscribing the ring visible in the image.
The criterion used to evaluate the match is simply the pixel-to-pixel
deviation between the two images, averaged over this region of
interest:

\begin{equation}\label{equ:sigma}
\sigma(C_{i},DC_{k})=\sqrt{<(C_{i}-DC_{k})^2>}
\end{equation}

The operation is repeated for all images $(C_i)_{i=1}^n$ of the CLOWFS
sequence. The matching short exposure science camera images (DS$_{k}$)
of the dictionary are then co-added to form an estimate of the
residuals in the long exposure coronagraphic image that can be
attributed to varying low-order aberrations during the exposure on the
science target. This final image is simply subtracted from the long
exposure to calibrate these pointing error residuals.

\section{Laboratory Demonstration}\label{sec:exp}

\begin{figure*}[htb!]
\centerline{\includegraphics[scale=0.5]{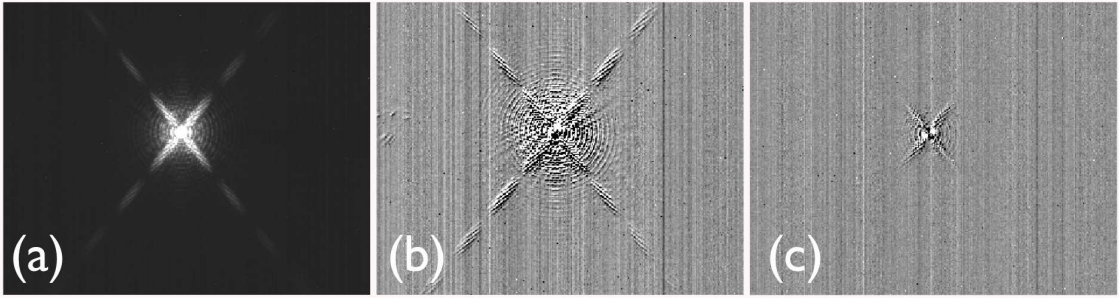}}
\caption{a) Long-exposure ``science camera'' image, non-linear intensity
  scale. Airy rings and diffraction spikes due to the peculiar Subaru
  Telescope pupil are clearly visible.
  b) Image after standard subtraction of a PSF without using the
  proposed CLOWFS image selection procedure.
  c) Image after CLOWFS post-processing calibration. Panels b and c
  use the same intensity scale.}
\label{fig:sci_img}
\end{figure*}

The experimental demonstration of the proposed post-processing
technique was performed using the Subaru Coronagraphic Extreme-AO
(SCExAO) bench in the Subaru laboratory
\citep{2009SPIE.7440E..20M}. Among extreme-AO systems, SCExAO
specializes in the high-contrast characterization of the innermost
($<0.2$ arc sec.) surrounding of stars, and in that scope, implements
a high-performance PIAA-based \citep{2003A&A...404..379G} coronagraph
that takes full benefit of the angular resolution of the 8-meter
Subaru Telescope. For our test, the special optics described by
\citet{2009PASP..121.1232L} were taken out of the beam and SCExAO was
reduced to a conventional near-IR Lyot-coronagraph without a Lyot-stop
\citep[see][for an introduction to coronagraphy with apodized
pupils]{2003EAS.....8...79A}. SCExAO is operated at IR wavelength
(H-band), and uses two identical uncooled fast readout InGaAs cameras
(Xenics model XS-1.7-320).

The light source is a $\lambda=1.55\mu m$ laser diode, illuminating a
17-mm pupil mask emulating the Subaru Telescope pupil, producing the
characteristic spikes in all the ``science'' images seen in this
paper. The focal-plane mask is 85 $\mu m$ in radius, which given the
f-number of the beam gives a $4.75$ $\lambda/D$ inner working angle
(IWA). Note that when the PIAA is inserted in the beam, the focal
plane scale is altered \citep{2005ApJ...622..744G} and the IWA becomes
$1.5 \lambda/D$.

For this experiment, a 25 minute sequence of 15000 pairs of 1
millisecond exposure images were acquired with the CLOWFS and
coronagraphic cameras on the SCExAO testbed at a $\sim 10$ Hz frame
rate. \footnote{During recent engineering observations with SCExAO,
the CLOWFS loop was successfully closed, using 5-10 ms exposures at an
improved 50 Hz frame rate.}
A series of 150 contiguous pairs of images in the middle of this
sequence were isolated to simulate a long-exposure acquisition. The
science images were dark-subtracted and co-added to simulate a single
long-exposure coronagraphic image. The corresponding sequence of raw
CLOWFS images was simply stored (they form the (C$_{i})_{i=0}^{149}$ list of
images). The other 14850 pairs of images, labeled
DC$_{k}$ and DS$_{k}$ were combined to form the dictionary.

The approach is illustrated by Fig. \ref{fig:sci_img}. The simulated
Science frame (i.e. co-addition of the 150 selected images) is located
in panel (a), while other panels compare the effect of two calibration
procedures: standard PSF subtraction, as explained in Section
\ref{Sec:eff} in panel (b) and the proposed CLOWFS post-processing
procedure in panel (c).

\section{Results}\label{Sec:results}

\subsection{Best match for a CLOWFS image}

For small wavefront aberrations, \citet{2009ApJ...693...75G} showed
that the CLOWFS image is a linear function of the low-order modes to
be measured, typically (but not restricted to) pointing, focus and
astigmatism.
The approach proposed here only uses the average pixel-to-pixel
deviation between two CLOWFS images, reducing the contribution of all
these modes to a single scalar, and does not attempt at separating the
different contributions.

\begin{figure}[htb!]
\centerline{\includegraphics[scale=0.4]{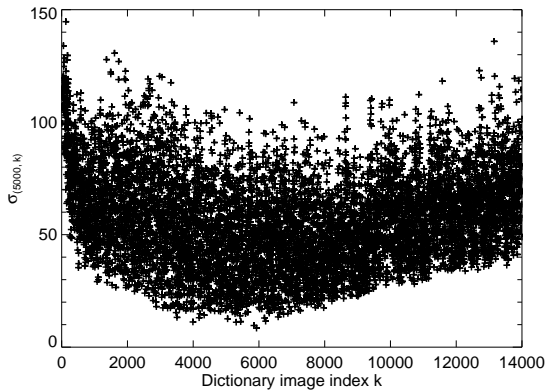}}
\caption{
  Evolution of the mean deviation
  $\sigma=\sqrt{<(DC_{5000}-DC_{k})^2>}$ over the
  25-minute acquisition sequence of the dictionary, using image
  index 5000 as reference. In addition to the slow trend that moves
  the average signal measured over the course of the sequence, the
  effect of the source of vibration used on the bench is obvious as a
  most of the recorded range of signal is explored in less than a
  minute.
}
\label{chi_all}
\end{figure}

To assess the extent of the coverage of these low-order modes by the
dictionary, one can arbitrarily pick one image as reference, and
calculate the deviation of each image of the sequence, relative to
this reference.
Fig.~\ref{chi_all} illustrates one such experiment, using image 5000
as reference. The value of this reduced CLOWFS signal ranges from
$\sigma\sim$10: the level readout noise of the detector used for this
experiment; to $\sigma\sim$140, with a mean at $\sim$55. Note for
reference, that in this configuration, a pure pointing error of
0.4$\lambda/D$ induces a reduced CLOWFS signal $\sigma\sim$110.

The 25-minute sequence of signal was acquired with a strong source of
vibration (an electric water pump) bolted to the bench, next to the
focal plane. In addition to the low trend visible on
Fig.~\ref{chi_all}, the fast vibration makes that CLOWFS quickly
explores the full range of aberrations covered during this experiment.
While building the dictionary, it is important to cover a sufficiently
wide amplitude of low-order modes signal, as in the post-processing
step, it improves the odds for the matching algorithm to find a
dictionary image that matches any CLOWFS image.

\begin{figure}[hb!]
\centerline{\includegraphics[scale=0.43]{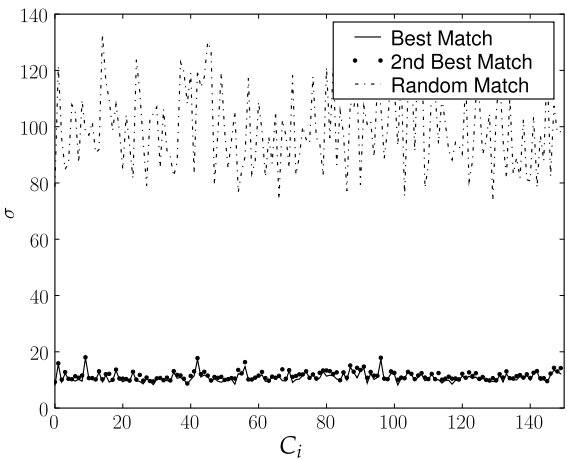}}
\caption{
  Mean deviation $\sigma$ in between each image $C_{i}$ of the CLOWFS
  sequence and its respective best and second best fit (full line and
  dots).
  The dash-dotted line shows the same deviation, this time between
  $C_{i}$ and a random CLOWFS image, averaged over 200 experiments.
  }
\label{sig_min}
\end{figure}

Fig.~\ref{sig_min}, shows as a solid line the value of the mean
deviation $\sigma$ between each image of the CLOWFS sequence its best
dictionary match. Hardly differentiable from the solid line is a
dotted line, showing the deviation with the 2$^{nd}$ best match. For
these two virtually identical cases, the deviation is dominated by the
detector readout noise ($\sim$10 ADU per pixel).
A randomly chosen dictionary image (dash-dotted line) exhibits a mean
deviation in average ten times as large.

\subsection{Calibration of coronagraphic leaks}
\label{Sec:eff}

Fig.~\ref{fig:sci_img} illustrates the significant improvement that
the calibration of low-order aberration induced coronagraphic leaks
with CLOWFS offers (c) over a more standard PSF subtraction (b).

The standard PSF subtraction shown in (b) was derived from the data
without knowledge of the CLOWFS frames.
To simulate the standard acquisition of a PSF for calibration in a
fair manner (in practice, this implies to find a single star of
comparable brightness and spectral type, located at a comparable
elevation), we randomly selected a number of images corresponding to
the total science exposure duration from the science side of the
dictionary and co-added them to form a second long-exposure frame
simulation. Fig.~\ref{fig:sci_img} (b) shows the residuals after the
PSF subtraction. While the outer part of the image is dominated by
detector readout, the scientifically valuable central part of the
field is dominated by residual speckle noise.

To produce the calibrated PSF shown in (c), the science images
corresponding to indices selected by the MMA procedure are co-added to
form what should be a very good PSF for the calibration of a
coronagraphic image.
The residual speckle level in this calibrated PSF is much lower than
for the standard PSF subtraction.

\begin{figure}[htb!]
\centerline{\includegraphics[scale=0.4]{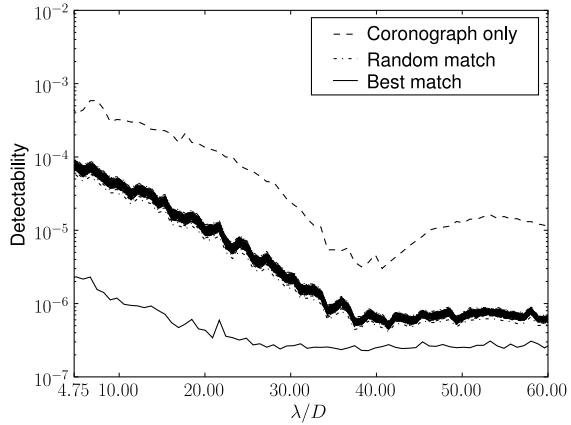}}
\caption{
  Standard deviation of the image profile as a function of angular
  separation. The figure compares three types of profiles: the raw
  science image (dashed line), the best-fit cleaned science image (solid
  line) using the proposed post-processing procedure and 200
  random-match cleaned science images (dot-dashed line) simulating the
  standard PSF subtraction.
}\label{radial}
\end{figure}

While only one example is provided in Fig.~\ref{fig:sci_img}, a total
of 200 different PSFs were also simulated, in order to provide a
statistically relevant comparison between the CLOWFS and standard PSF
subtraction residuals.
Fig.~\ref{radial} shows together all the radial profiles of raw,
standard-PSF-subtracted and CLOWFS-PSF-subtracted images. Each of the
200 simulated PSFs offer an improvement of the contrast over the
entire field of view by a factor ten, in comparison with the raw
image. In comparison, the CLOWFS PSF improves the calibration by
almost two orders of magnitudes between the inner working angle of the
Lyot-coronagraph (4.75 $\lambda/D$) and $\sim$20$\lambda/D$, where the
contrast reaches the detector readout noise floor.
In practice, a standard PSF subtraction would require a change of the
pointing of the telescope to observe a calibration star. Such a change
will change the structure of static aberrations, introducing a
systematic error that will degrade actual performance of the PSF
subtraction in comparison with what is presented here. On sky,
CLOWFS-like calibrations are likely to prove even more advantageous.

\section{Conclusion}\label{sec:discussion}

The results presented in Sec~\ref{Sec:results} demonstrate that, used
in post-processing, CLOWFS improves the sensitivity of a coronagraph
by a factor 40, in comparison to more conventional -yet idealized-
type of calibration.
The method described here for this proof of concept is yet still
quite rudimentary, and could benefit from several improvements.

Instead of relying on an exhaustive search by a MMA-like algorithm on
a massive database, necessarily containing redundant information,
the dictionary could be simplified using approaches like Principal
Component Analysis, in which CLOWFS images would be projected onto a
smaller sub-set of CLOWFS modes.
In addition, an extended knowledge of several experimental parameters,
such as the telescope elevation, star color or magnitude, could be
implemented in the dictionary so as to improve on the actual
calibration.

The approach is nevertheless powerful as it requires neither modelling
of the coronagraph, nor assumptions such as linearity of the response
in order to work. The absence of such requirements makes it extremely
robust, and able to handle multiple situations and types of
coronagraphic leaks.

The Subaru Coronagraphic Extreme AO Project implements a CLOWFS, used
in close-loop. The system architecture however allows for a simple
implementation of the CLOWFS post-processing, with no impact on the
hardware, to calibrate residual low-order modes induced coronagraphic
leaks.
CLOWFS on SCExAO currently generates a steady 4 Mbits/s data
stream. While not prohibitively large for an experiment, the storage
of entire nights of CLOWFS data will quickly prove impractical as
SCExAO enters a more aggressive observing phase. Eventually,
the calibration of coronagraphic leaks using CLOWFS will require the
matching to be performed on-the-fly, so that only the final
coronagraphic leak image is actually saved.

Because it makes no assumption on targets and the current state of the
optical bench, CLOWFS used both for live and post-processing is likely
to play a significant part to the success of SCExAO, focused on the
detection of planetary companions at small angular separation.

\acknowledgments

Fr\'ed\'eric Vogt thanks Brian Elms for sharing his experience and the
machining of parts used for this project.

{\it Facilities:} \facility{Subaru (SCExAO)}.

\end{document}